\documentclass[mathpazo]{cicp}

\usepackage{amsmath}
\usepackage{graphicx}
\usepackage{dcolumn}
\usepackage{bm}
\usepackage{color}
\usepackage{epsfig}
\usepackage{mathrsfs}

\begin{document}
\title{Understanding depletion induced like-charge attraction from self-consistent field model}

 \author[P. Liu, M. Ma, Z. Xu]{Pei Liu\affil{1},
       Manman Ma\affil{1}, and Zhenli Xu\affil{1,2}\comma\corrauth}
 \address{\affilnum{1}\ School of Mathematical Sciences and Institute of Natural Sciences, Shanghai Jiao Tong University, Shanghai 200240, P.R. China. \\
           \affilnum{2}\ MoE Key Lab of Scientific and Engineering Computing, Shanghai Jiao Tong University, Shanghai 200240, P.R. China.}
 \emails{{\tt hgliupei1990@sjtu.edu.cn} (P.~Liu), {\tt mmm@sjtu.edu.cn} (M.~Ma),
          {\tt xuzl@sjtu.edu.cn} (Z.~Xu)}

\begin{abstract}
The interaction force between likely charged particles/surfaces is usually repulsive due to the Coulomb interaction. However, the counterintuitive like-charge attraction in electrolytes has been frequently observed in experiments, which has been theoretically debated for a long time. It is widely known that the mean field Poisson-Boltzmann theory cannot explain or predict this anomalous feature since it ignores many-body properties. In this paper, we develop efficient algorithm and perform the force calculation between two interfaces using a set of self-consistent equations which properly takes into account the electrostatic correlation and the dielectric-boundary effects. By solving the equations and calculating the pressure with the Debye-charging process, we show that the self-consistent equations could be used to study the attraction between like-charge surfaces from weak-coupling to mediate-coupling regime, and that the attraction is due to the electrostatics-driven entropic force which is significantly enhanced by the dielectric depletion of mobile ions. A systematic investigation shows that the interaction forces can be tuned by material permittivity, ionic size and valence, and salt concentration, and that the like-charge attraction exists only for specific regime of these parameters.
\end{abstract}

\pac{82.45.Un, 
64.70.pv,  
82.60.Lf 
}

\maketitle

\section{Introduction}
\label{intro}
The anomalous attraction between likely charged particles has been widely discussed during the past decades  \cite{FPP+:RMP:2010} since the experimental observation in colloidal \cite{Dishon:Langmuir:2009} and biological systems \cite{GBP+:PT:2000,ALW+:PNASU:2003}. The like-charge attraction (LCA) at strong-coupling regime (high surface charge, multivalent counterions or low temperature) has been reported by particle-based simulations  and  theoretical studies such as density functional theory and strong-coupling theory \cite{LL:PRL:99,DTBL:PA:1999,Linse:JCP:00,NN:EPJE:04,MHK:PRL:2000,JKN+:PRL:2008,Messina:JPCM:2009,NKFP:JCP:13}. Colloidal suspensions in monovalent electrolytes are usually considered as systems in weak-coupling regime at which the mean-field Poisson-Boltzmann (PB) theory is generally useful. The forces between charged surfaces are usually modelled through the Derjaguin-Landau-Verwey-Overbeek (DLVO) theory, which combines the linearized PB equation with the Van der Waals interaction. However, the PB theory always predicts repulsive interaction between like charges \cite{Neu:PRL:1999,Trizac:PRE:00}, and the DLVO theory overestimates the repulsion or underestimates the attraction \cite{HCL:JCP:2008}, in contrast to the results of LCA experiments for colloids \cite{LG:N:1997,krishnan2008a},  which demonstrates many-body effects are still significantly important at weak-coupling conditions and can not be ignored in order to understand the LCA phenomenon systematically.

For two colloidal particles with a small separation, the depletion force induced by the entropic repulsion of smaller mobile ions in the solvent could result in an effective inter-colloid attraction. The excluded volumes of small particles lead to a depletion layer. When the separation between surfaces is at the range of the particle diameter, they are attractive due to this depletion interaction \cite{Asakura:JCP:1954,Pincus:PRE:1999}. In electrolytes, this attractive distance can be much larger, comparable with the Debye screening length due to the long-range nature of electrostatic interaction \cite{CL:COCIS:2015}. In this sense, the attractive Van der Waals and the Casimir force \cite{Casimir:PKNAW:1948,Podgornik:EPL:2015,Samaj:JSM:2004} can be ignored in the LCA analysis. In the presence of inhomogeneous dielectric permittivity (whose properties have attracted wide recent interest \cite{WM:JPCB:2010,DL:JPCM:2012a,JSC:JCP:13,ZD:PNAS:13,BL:PRL:2014,Derbenev2016,Emelyanenko2016}), ions are repelled from the low dielectric surfaces by their image charges, which has been shown in particle-based simulations \cite{BH:PRE:1994,Gan:JCP2015a}. When the separation of surfaces becomes narrower, the repulsion becomes stronger, resulting in lower ionic concentration between the surfaces, and thus leading to an entropic driven attraction. The electrical field can also influence the solvent alignment and dielectric property, which is considered to be important for high ionic concentration or at the strong coupling region. This solvent polarization effect could be modelled through the Langevin theory to yield a field-dependent dielectric coefficient \cite{B:JCP:1951,B:JCP:1955,Azuara:BJ:2008}. Recently, a continuum theory with a set of self-consistent equations including fluctuation effects of ions is introduced as a variational approach with general Gaussian ansatz \cite{NO:EPJE:2003,Wang:PRE:2010}, which have been generalized to take into account the image charge effects on attractive forces between neutral plates and in weak coupling region \cite{WangRui:JCP:13}. In the theory, the self energy of mobile ions is used as a correction to the mean potential in the PB theory, leading to a more accurate approximation of the potential of mean force (PMF) in the Boltzmann distribution, which consists of contributions from both the local ionic correlation and the ion-interface interaction energy. Although the LCA phenomenon has been predicted, it is less understood what  is the effects of salt property as well as the charged surface.

In this paper, we study the interaction between plates in electrolytes by the self-consistent field model developed in \cite{XML:PRE:2014,MX:JCP:14} which includes a modified treatment of the self energy. Through the modification, we include the ionic size effect to avoid the ion collapse in the original theory and are able to study systems from weak-coupling to mediate-coupling regimes and with a inhomogeneous dielectric coefficient beyond the point-charge models. We present a systematic analysis for effects of the surface charge density, the bulk salt density, and salt species on the depletion-induced LCA. It is shown that for symmetric electrolytes the interaction force between plates is repulsive when the surface charge density is high or when the salt concentration is low. The force becomes attractive at the opposite regimes, for which the strength will be stronger when the charge density is decreased or the salt density is increased.

\section{Self-consistent field model}

Consider an equilibrium charged system with $N$ species of mobile ions in solution. Given the PMF of the $i$th species, $w_i(\mathbf{r})$, the ion distribution can be determined by the Boltzmann distribution,
\begin{equation}
\displaystyle c_i(\mathbf{r}) = c_i^b e^ {-\beta w_i(\mathbf{r})},
\end{equation}
where $c_i^b$ is the bulk concentration, and $\beta= 1/{k_B T}$ is the inverse thermal energy with $k_B$ the Boltzmann constant and $T$ the absolute temperature. The PMF describes the free energy change by moving the ion from the bulk region into the current position, which is often given by particle simulations or advanced theory in statistical physics. If the PMF is approximated by the mean potential energy $w_i=z_ie \Phi$, with $\Phi$ being determined by the Poisson equation,
\begin{equation}
-\varepsilon_0\nabla\cdot\varepsilon \nabla\Phi = \sum_{i=1}^N z_ie c_i,
\end{equation}
where $\varepsilon_0$ is the vacuum dielectric constant, $\varepsilon$ is the relative permittivity of the media, and $z_i$ is the valence, we obtain the PB equation since $c_i$ is now an explicit function of $\Phi$.

To properly take into account the dielectric polarization and the correlation effects of mobile ions, one should study the self energy or the intrinsic chemical potential of a test ion, which is considered as a correction term to the mean potential to better approximate the PMF \cite{AM:CJU:1986,podgornik1989jcp,NO:EPJE:2003,BAA:JCP:2012,HL:SM:2008, BMP:PRE:2010, WangRui:JCP:13,XML:PRE:2014,LX:PRE:2014,ma2016}. In our model, we treat an ion as an ion-inaccessible sphere of radius $a_i$ with the point charge $z_i$ at the center. This allows us to deal with variable dielectric media where the ionic Born energy, which strongly depends on the ion radius, is not constant and thus cannot be discarded. For a homogeneous system, the self energy of an ion can be described by the exact solution of the Debye-H\"uckel equation \cite{Levin:RPP:2002},
\begin{equation}
U_i = \frac{z_i^2 e^2 }{2}u_i= \frac{z_i^2 e^2 \kappa}{8\pi \varepsilon_0 \varepsilon } \frac{1}{1 + \kappa a_i}, \label{lda}
\end{equation}
where $\kappa$ is the inverse Debye length of the electrolyte.
In the vicinity of a surface, the ionic distribution and the dielectric permittivity are space-dependent, and a direct use of Eq. \eqref{lda} by replacing $\kappa$
by the space-dependent quantity cannot account for the nonlocal property of the electrostatic correlation. We assume that the ion size is small and that the presence of a test ion does not influence  the ionic distribution in the electrolyte except for the region occupied by the ion due to the finite excluded volume. The self energy of the $i$th ion species is then formulated through the generalized Debye-H\"uckel equation \cite{XML:PRE:2014},
\begin{equation}
\begin{cases}
\displaystyle -\varepsilon_0 \nabla \cdot \varepsilon_i \nabla G_i(\mathbf{r},\mathbf{r}') + 2I_i  G_i(\mathbf{r},\mathbf{r}') = \delta(\mathbf{r},\mathbf{r}'), \label{green1}\\
\displaystyle u_i(\mathbf{r})= \lim_{\mathbf{r}'\to \mathbf{r}}[G_i(\mathbf{r},\mathbf{r}')-G_0(\mathbf{r},\mathbf{r}')],
\end{cases}
\end{equation}
where the dielectric permittivity and the ionic strength locally depend on the site of the test ion,
\begin{equation}
\varepsilon_i(\mathbf{r},\mathbf{r}')=\begin{cases}
1 , \ \ \ \ \hbox{if}~~|\mathbf{r}-\mathbf{r}' |\leq a_i,\\
\varepsilon(\mathbf{r}) , \ \ \ \hbox{otherwise},
\end{cases}
\end{equation}
\begin{equation}
\displaystyle I_i(\mathbf{r},\mathbf{r}') =\begin{cases}
0, \ \ \ \ \hbox{if}~~ |\mathbf{r}-\mathbf{r}' |\leq a_i,\\
I(\mathbf{r}), \ \ \ \hbox{otherwise},
\end{cases}
\end{equation}
with
\begin{equation}I(\mathbf{r})=\frac{1}{2}\sum_i\beta z_i^2 e^2 c_i(\mathbf{r})\end{equation}
being the mean ionic strength.
The function $G_0$ in Eq. \eqref{green1} is the Green's function in free space, satisfying
\begin{equation}-\varepsilon_0  \nabla^2 G_0 = \delta(\mathbf{r},\mathbf{r}'),
\end{equation}
which is used to eliminate the singularity in $G_i$.

When $\varepsilon$ and $I$ are constant, the solution $u_i$ reduces to the Debye-H\"uckel theory \eqref{lda}. The equation for $G_i$, Eq. \eqref{green1}, being considered as the generalization of the Debye-H\"uckel equation is in the sense that the inverse Debye length depends on the space, $\kappa(\mathbf{r})=\sqrt{I(\mathbf{r})/2\pi\varepsilon_0\varepsilon_{W}}$, where $\varepsilon_\mathrm{W}$ is relative dielectric constant of water. With the expression for the self energy, we then obtain a modified PB equation,
\begin{equation}
-\varepsilon_0\nabla\cdot\varepsilon\nabla\Phi = \sum_{i=1}^N z_ie \widetilde{c}_i^b \exp\left[-\beta \left(z_ie \Phi+\frac{ z_i^2 e^2}{2}u_i\right)\right], \label{mpb}
\end{equation}
where $\widetilde{c}_i^b=c_i^b \exp \left( \beta z_i^2 e^2 u_i^b/2\right)$ and $u_i^b$ is the bulk value of $u_i$.

So far, we have described a continuum model for inhomogeneous electrolytes through a set of self-consistent field (SCF) Eq. \eqref{green1}-\eqref{mpb}. This set of equations include the long-range correlation by the solution of the generalized Debye-H\"uckel equation. However, the model may make imprecise prediction for dense electrolytes or strongly-correlated systems, because the hard-core correlation is neglected. In spite of this,  it was reported that the SCF model can capture correlation phenomena well for electrolyte systems up to the coupling parameter $\Xi\sim50$ \cite{MX:JCP:14}, which is certainly in the regime of strong coupling.

\section{Method}
\subsection{Physical setup}
\begin{figure}[t]
\begin{center}
\includegraphics[width=0.6\textwidth]{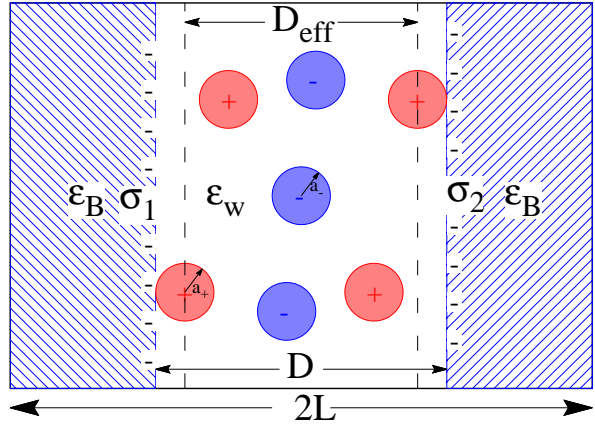}
\caption{ A schematic description of electrolytes between two likely charged planes of separation $D$. Ions are of finite sizes and are inaccessible to a zone of thickness $a_+$ from the surfaces. If the dielectric constant outside the electrolyte is much smaller than that of the solvent, a repulsive self energy will be acted on each mobile ion. } \label{schem}
\end{center}
\end{figure}

We compute the interaction forces between two charged surfaces in a system schematically shown in Fig. \ref{schem}. The two parallel planes are with surface charge densities $\sigma_1$ and $\sigma_2$, respectively. The planes are orthogonal to the $x$ axis, locating at $x=\pm D/2$. An electrolyte is in between the two planes, and its relative dielectric permittivity is $\varepsilon_\mathrm{W}$. The outside space is filled with dielectric media characterized by the relative dielectric constant $\varepsilon_B$. We assume negative surface charges on two planes, and study electrolytes with two species of ions (counterions are positive ions). Since mobile ions have finite sizes, there is a zone, i.e. the Stern layer, of the ion-radius thickness inaccessible by mobile ions near each charged plane.

Due to the planar geometries, the ionic density is invariant along the $y-z$ plane parallel to the surfaces. The modified PB equation \eqref{mpb} is one-dimensional, and can be iteratively solved by a finite-difference method, giving the self energy. In the calculation, Neumann boundary conditions $\frac{\partial \Phi}{\partial x}=0$ are applied on two ends of interval $[-L, L]$ with $L>D/2$ big enough to remove artifacts from the boundary. The surface charge placed at $\pm D/2$ is approximated by the Kronecker delta with a coefficient inversely proportional to the mesh size.

\subsection{Numerical method}

The generalized Debye-H\"uckel equation \eqref{green1} is high dimensional and hard to solve. In order to develop an efficient numerical method, we first employ an asymptotic treatment \cite{MX:JCP:14}, which approximately separates the self energy as the sum of two contributions,
\begin{equation}
u_i=u_i^{Born}+u_i^{Coul}, \label{separation}
\end{equation}
under the assumptions that the ionic radius $a_i$ can be considered as a small parameter and that the ionic strength is slowly varying.
The first term in Eq. \eqref{separation} is the Born solvation energy\cite{Born:ZP:1920,Wang:PRE:2010},
\begin{equation}
u_i^{Born} = \frac{1}{4\pi \varepsilon_0 a_i}\left(\frac{1}{\varepsilon}-1\right),
\label{born}
\end{equation}
The second term containing the contribution from  the long-range Coulomb correlation is expressed as,
\begin{equation}
u_i^{Coul} = \frac{u}{1 + a_i\kappa_b},
\label{coulomb}
\end{equation}
where $\kappa_b$ is the inverse Debye length in the bulk. The value of $u$ is determined by the following equations,
\begin{equation}
\begin{cases}
\displaystyle -\varepsilon_0 \nabla \cdot \varepsilon \nabla G(\mathbf{r},\mathbf{r}') + 2I G(\mathbf{r},\mathbf{r}') = \delta(\mathbf{r},\mathbf{r}'), \label{green}\\
\displaystyle u(\mathbf{r})= \lim_{\mathbf{r}'\to \mathbf{r}}[G(\mathbf{r},\mathbf{r}')-G_0(\mathbf{r},\mathbf{r}')].
\end{cases}
\end{equation}
This formulation \eqref{coulomb} is similar to the description of the intrinsic chemical potential in Ref. \cite{Roux:BJ:1997}. The form of $-\kappa_b/(1+\kappa_b a)$ as the correlation energy was also established by the means of integral equation theory \cite{McQuarrie:JCP:1976}. For homogeneous system $I=\kappa_b^2\varepsilon_0\varepsilon_\mathrm{W}/2$ and $\varepsilon(\mathbf{r})=\varepsilon_{\mathrm{W}}$, it is easy to see $u= -\kappa_b/4\pi\varepsilon_0\varepsilon_\mathrm{W}$. However, for systems with strong inhomogeneity, such as those in our results, $u$ can be largely different from $-\kappa_b/4\pi\varepsilon_0\varepsilon$. For such systems, we find our asymptotic expression \eqref{coulomb} works well. In the previous work \cite{MX:JCP:14}, another expression $u_i^{Coul} \approx u/(1 - 4\pi \varepsilon_0 \varepsilon  a_i u)$ was used, which also works if $u$ does not become positive. In this study, $u$ could become a large positive number near charged surfaces, thus this is essential to use expression \eqref{coulomb} in order to avoid the zero-denominator instability.

Eq. \eqref{green} can be derived through field theoretical approach and has been studied as the self energy of a point charge. The advantage of Eq. \eqref{green} compared to Eq. \eqref{green1} is that it removes the multiple length scales in the model and can be numerically computed efficiently \cite{XuMaggs:JCP:14}. Then the equation can be better understood if we write it into the form: $AG=\delta$ where $A = -\varepsilon_0 \nabla \cdot \varepsilon \nabla + 2I$ is an operator which can be viewed as a matrix and $\delta$ can be viewed as the identity matrix. Since we are only interested in the diagonal entries of $G$, the selected inversion algorithm \cite{LYM+:ATMS:2011} could be used and it is very efficient.
Considering the geometric symmetry, we employ the Fourier transform in the $y-z$ coordinates to the Green's function equation,
\begin{equation}
\left[-\varepsilon_0  \frac{\partial}{\partial x} \varepsilon \frac{\partial}{\partial x}+ \varepsilon_0\varepsilon\omega^2 +2I(x)
\right] \widehat{G}(\omega; x,x') = \frac{1}{2\pi}\delta(x-x').
\end{equation}
In domain $x\in[-L, L]$, we could discrete the equation by finite-difference method with periodic boundary conditions, which is in the form of: $A_n\widehat{G}=I_n$, where $A_n$ is a $n \times n$ symmetric sparse matrix and $I_n$ is the identity matrix. Then the diagonal part of $G$ can be solved by the selected inversion method, whose computational cost is at the order of Cholesky factorization. The divergent part of the self-Green's function is eliminated by numerically solving $\widehat{G}_0(\omega; x,x)$ by the same scheme, then $\widehat{G}-\widehat{G}_0$ is convergent with the refinement of the mesh size.  The self energy is then determined by the inverse Fourier transform \cite{XuMaggs:JCP:14}.

We should also notice that, as $G_0 \sim 1/|\mathbf{r}-\mathbf{r}'|$ decays very slowly, we can hardly guarantee the precision of $G_0$ without choosing relatively large computational domain. To avoid this inconvenience, instead of solving $
-\varepsilon_0  \nabla^2 G_0 = \delta(\mathbf{r},\mathbf{r}')$, we solve $
- \nabla^2 \widetilde{G}_0 + k^2 \widetilde{G}_0= \delta(\mathbf{r},\mathbf{r}')/\varepsilon_0$ where $k>0$ is a constant and use the fact that $\lim_{\mathbf{r}'\to\mathbf{r}} (\widetilde{G}_0- G_0 )= -k/4\pi\varepsilon_0$.

Now we have described the method for solving the self energy of ions with given ionic distributions. With the numerically computed self energy,
Eq. \eqref{mpb} for the electric potential is discretized through the finite difference method of second order accuracy.
Since the ionic distributions are nonlinear functions of the electric potential, we employ the quasi-Newton method for the iteration
in solving the nonlinear PDEs. Then the updated ionic concentrations are given by Boltzmann's distribution and the whole system is solved
through the fixed-point iteration until the final self energy and electrical potential are self-consistent to the required precision
(error less than $10^{-8}$).
After solving the set of self-consistent equations, we obtain the equilibrium electrical potential, ionic concentrations as well as the self energy.

\subsection{Free energy and pressure}
In order to measure the interaction force between the two planes, we first calculate the grand potential per unit area as the function of the separation,
\begin{equation}
\displaystyle F(D) = \int_{-D/2}^{D/2}  \left[ \frac{\varepsilon_0\varepsilon (\partial_x \Phi)^2}{2}+ k_BT\sum_i c_i \left(\log \frac{c_i}{c_i^b} -1\right)+f_\mathrm{fl}\right]dx,
\label{energy}
\end{equation}
where $f_\mathrm{fl}$ is the contribution of fluctuation, given by the Debye charging process \cite{Bell:JCP:1968a,WangRui:JCP:13},
\begin{equation}
f_\mathrm{fl}=\sum_i z_i^2 e^2 c_i \int_0^1 \lambda \left[\frac{u_\lambda}{1 +  a_i \lambda \kappa_b} + \frac{1}{4\pi \varepsilon_0 a_i} \left(\frac{1}{\varepsilon}-1\right) \right] d\lambda,
\label{fl}
\end{equation}
and $u_\lambda$ is determined by,
\begin{equation}\left\{
\begin{array}{ll}
\displaystyle -\varepsilon_0 \nabla\cdot \varepsilon \nabla G_\lambda(\mathbf{r},\mathbf{r}') + 2\lambda^2  I  G_\lambda(\mathbf{r}, \mathbf{r}') = \delta(\mathbf{r}, \mathbf{r}'),\\
\displaystyle u_\lambda(\mathbf{r})= \lim_{\mathbf{r}'\to \mathbf{r}}[G_\lambda(\mathbf{r},\mathbf{r}')-G_0(\mathbf{r},\mathbf{r}')].
\end{array}
\right.
\label{ulambda}
\end{equation}
The fist two terms in Eq. \eqref{energy} represent the electrostatic and ideal-gas entropic contributions on the free energy.
The third term is the correlation free energy. It should be emphasized that the entropy of mobile ions should have additional contribution due to the finite size, e.g., by the modified fundamental measure theory \cite{Rosenfeld:PRL:1989,Roth:JPC:2002,Wu:JCP:2002}. For electrolytes without dense ionic particles, this contribution has minor effect.

The integral in Eq. \eqref{energy} is calculated through the trapezoidal rule, which has a second order of accuracy.
The main difficulty of evaluating the free energy and pressure is to compute the charging process \eqref{fl}. When $\lambda$ is small, due to the weak screening, the precision of $G_\lambda$  requires a large computational domain. Although no explicit solution is valid, we could not use the same technique as for $G_0$. It is noticed that the leading term of $G_\lambda - G_0$ is proportional to $\lambda$ when $\lambda$ is small. So, we can choose a relatively small number $\alpha$, inversely proportional to the computational domain $L$, which could guarantee that the accuracy of $G_\alpha$. If $\lambda<\alpha$, we use the linear interpolation of $G_0$ and $G_\alpha$ instead of solving \eqref{ulambda}. Thus, the integration in Eq. \eqref{fl} can be divided into two parts: (1) for small $\lambda$, the integration from 0 to $\alpha$ is integrated analytically ; (2) the integration from $\alpha$ to 1 is computed with Gauss quadrature. The error from both parts could be well controlled.

With the grand potential, the osmotic pressure per unit area is determined by
\begin{equation}\displaystyle P= - \partial_D F -P_\infty,
\end{equation}
where  $\displaystyle P_\infty=- \lim_{D \to \infty} \partial_D F$ is the bulk osmotic pressure.

\section{Results and discussion}

In this section, we present numerical solutions of the self-consistent equations for the modified PB theory and show different aspects of the LCA phenomenon. Without special statement, we take relative dielectric constants $\varepsilon_B=2.5$ and $\varepsilon_W=80.$ The differential equations are calculated in a very fine mesh (mesh size $\Delta x=0.01 nm$) in order to accurately approximate the solutions. All systems are at room temperature, i.e., $\ell_B=0.714 nm$.
Only electrolytes with two ion species are studied, we will use $\pm$ for subscript $i$ to distinguish cations and anions.

\begin{figure*}[t]
\begin{center}
\includegraphics[width=0.49\textwidth]{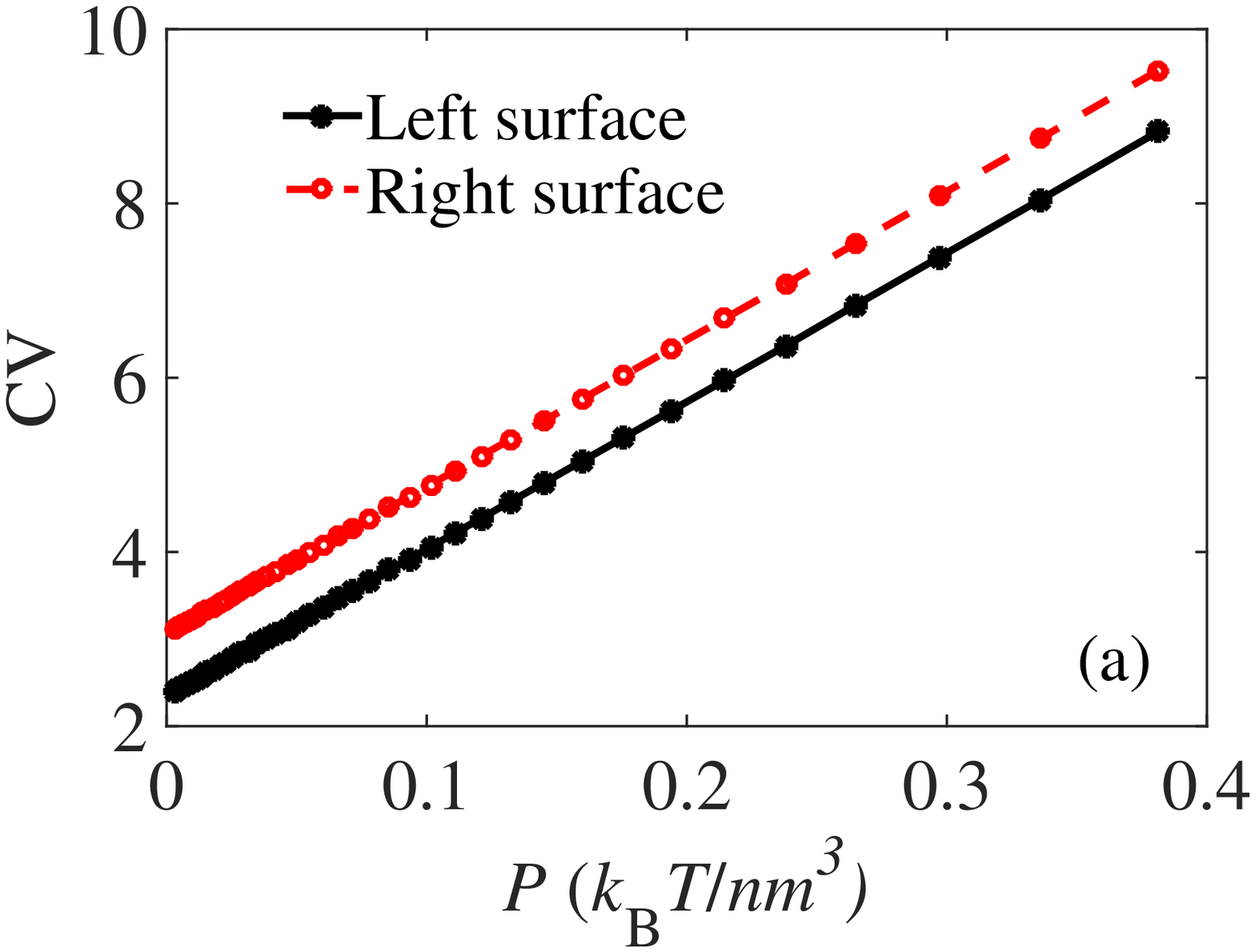}
\includegraphics[width=0.49\textwidth]{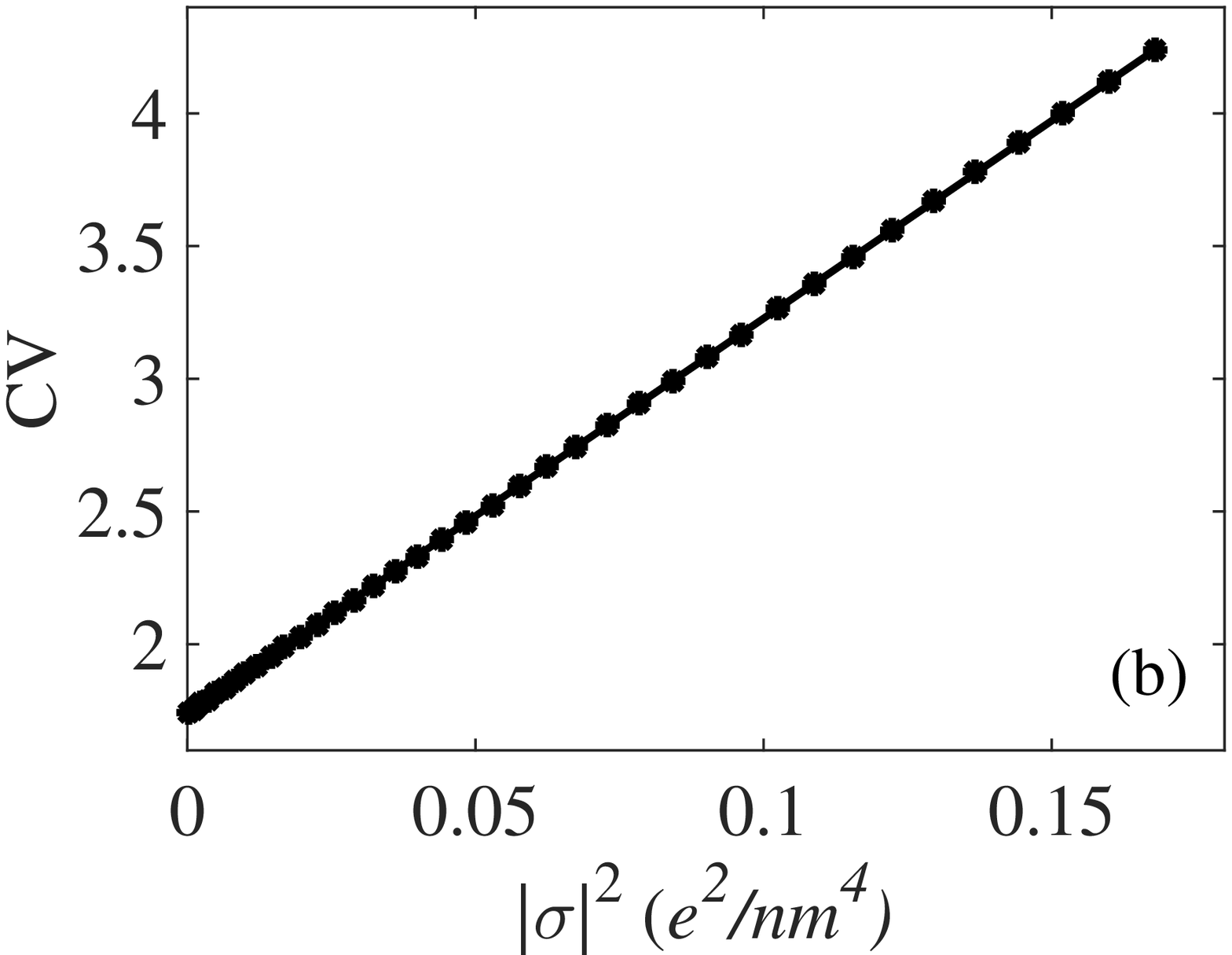}
\caption{Numerically calculated contact values.   (a) The profiles as function of pressure for given surface charge densities $\sigma=[-0.01,-0.1] e/nm^2$. In the calculation, $D$ increases from $0.6nm$ to $1.4nm$ such that $P$ and $CV$ are varied. (b) The profile as function of $\sigma^2$ for given pressure. In the calculation, the surface separation is fixed to be $3nm$ so that the system pressure can be viewed as a constant equal to the electrolyte bulk pressure. } \label{contact}
\end{center}
\end{figure*}

Before proceeding the calculation for the LCA, we demonstrate that the SCF model numerically satisfies the contact value theorem \cite{HM::2006}
and thus the model is reliable from this point. The theorem for monovalent electrolytes is an exact relation between the total
contact ionic concentration, the system pressure $P$ and the surface charge density $\sigma$,
\begin{equation}
k_B T \sum_i c_i(x_\mathrm{surf}) = P + \frac{2\pi \sigma^2}{\varepsilon_0\varepsilon},
\end{equation}
where $c_i(x_\mathrm{surf})$ is the ionic density of the $i$th species on the surface at $x_\mathrm{surf}$. the classical PB theory satisfies this relation
approximately by assuming the pressure is the ideal-gas pressure in the bulk. Recent study has
studied the contact value theorem for the SCF for the point-charge case \cite{FM:PRE:2016}.
Our model is more complicated because we cannot find explicit expressions for both the pressure and the contact
concentration. We use $0.1 M$ 1:1 electrolytes with $a_\pm=0.2 nm$ and the dielectric permittivity is 80 in whole space.
We calculate the normalized contact values numerically, $CV = \sum_i c_i(x_\mathrm{surf})/c_{ib}$, as functions
of $P$ for given $\sigma$ and $\sigma^2$ for given $P$, respectively. The two panels in Fig. \ref{contact} show
the linear relations for both cases, which demonstrate that the contact value theorem is fulfilled.
Compared to the classical PB theory, the SCF model presents more accurate results
since the approximation for the pressure with the ideal-gas pressure is not necessary.

\subsection{Symmetric electrolytes}

\begin{figure*}[htbp]
\begin{center}
\includegraphics[width=0.49\textwidth]{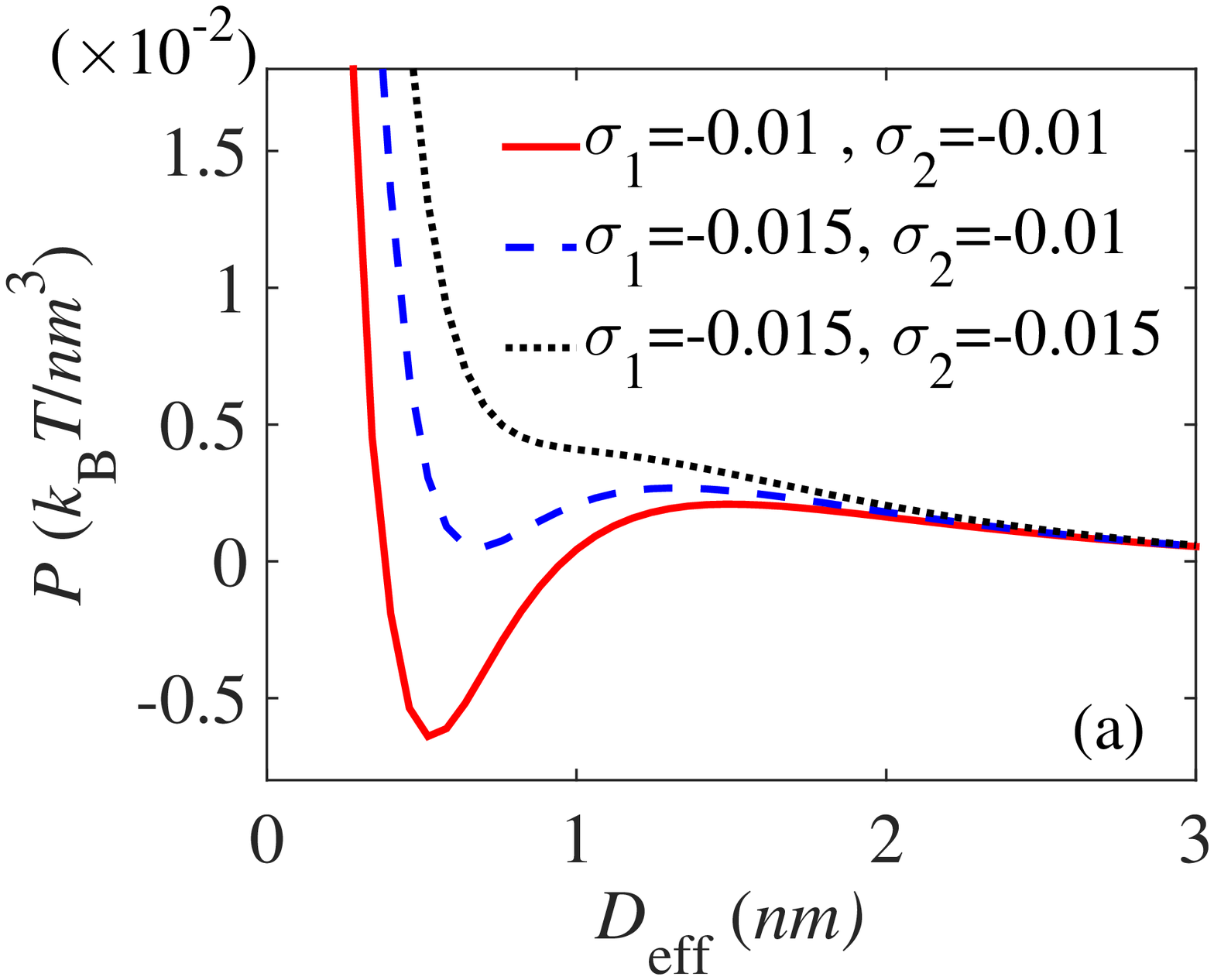}
\includegraphics[width=0.49\textwidth]{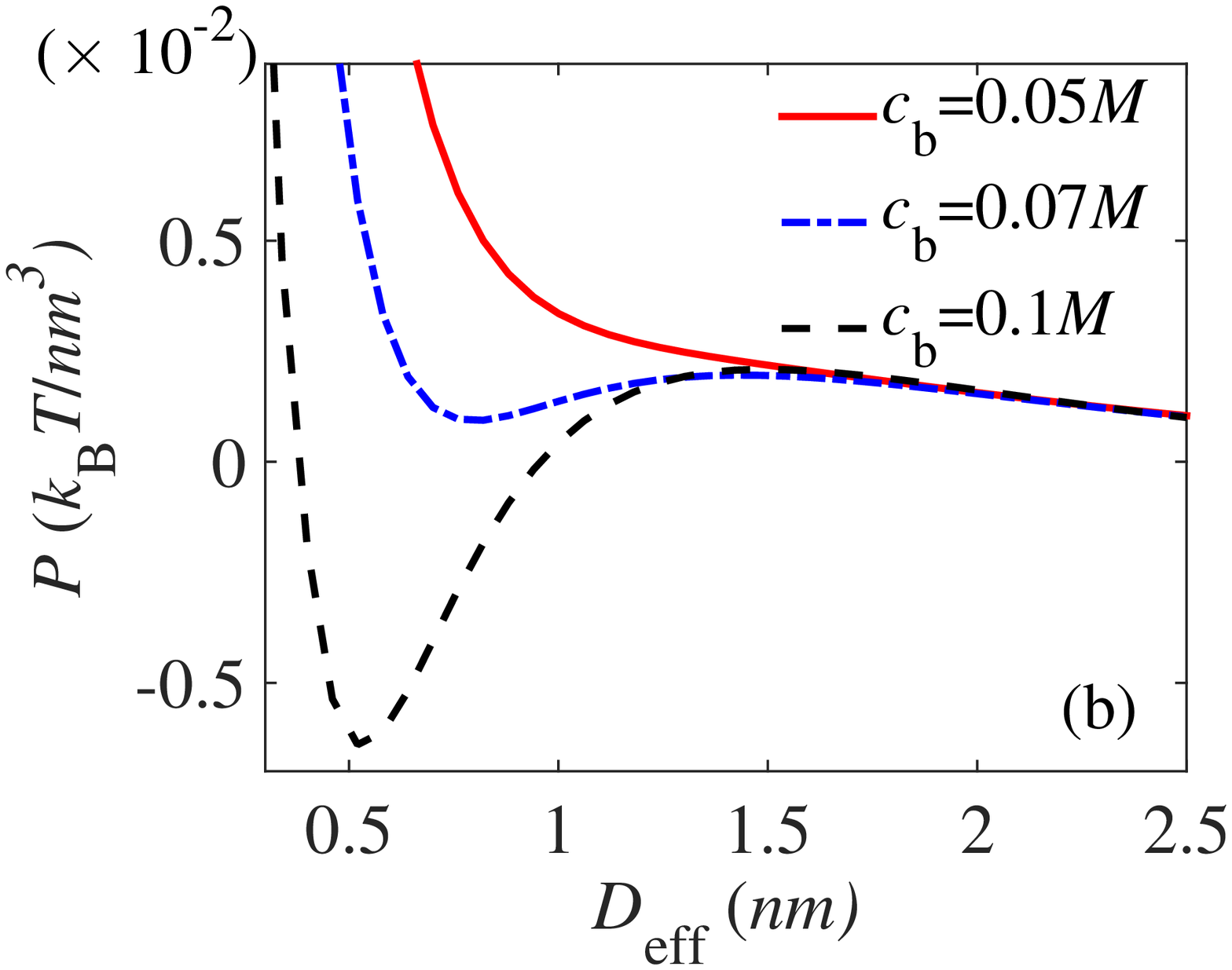}\\
\includegraphics[width=0.49\textwidth]{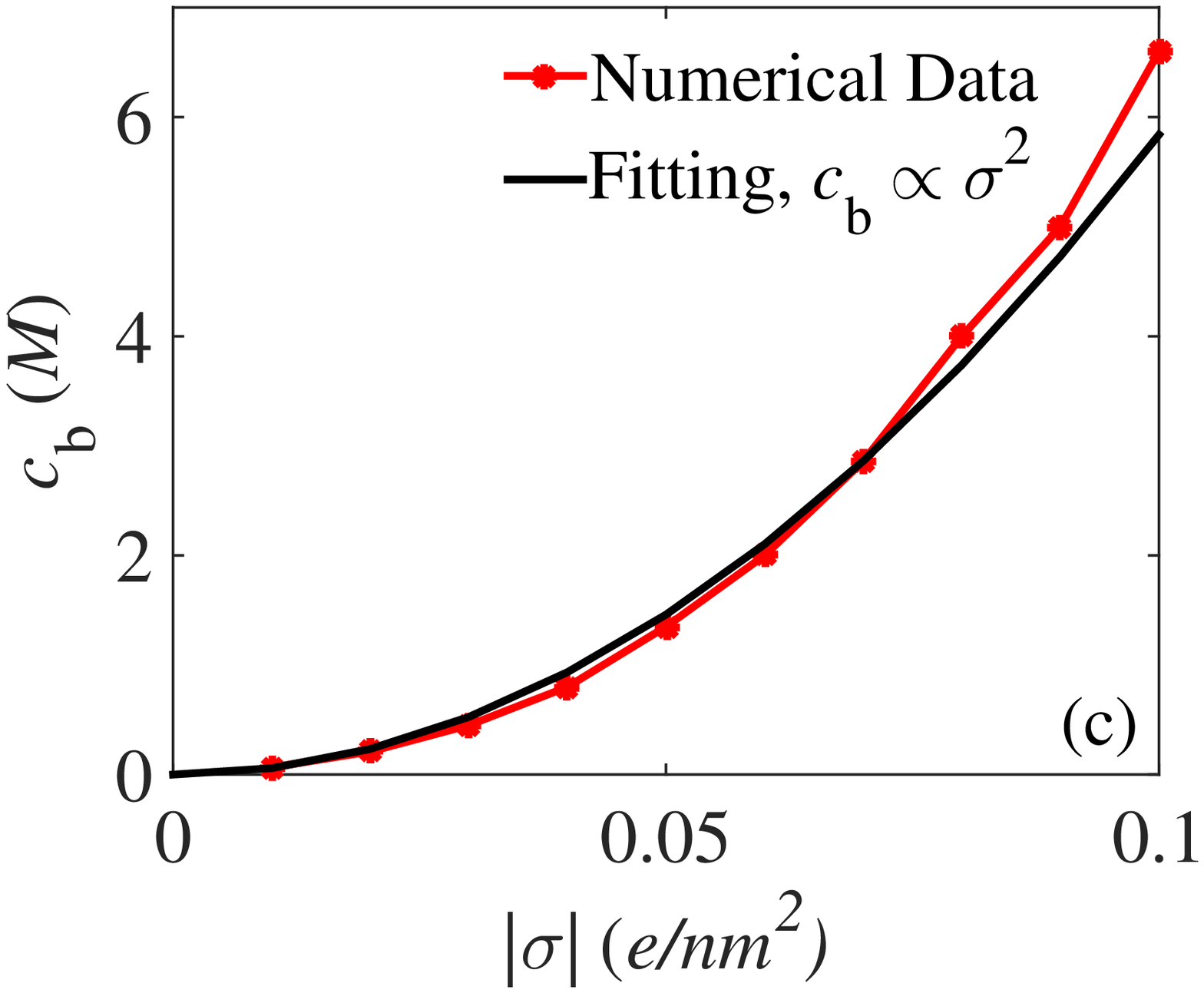}
\includegraphics[width=0.49\textwidth]{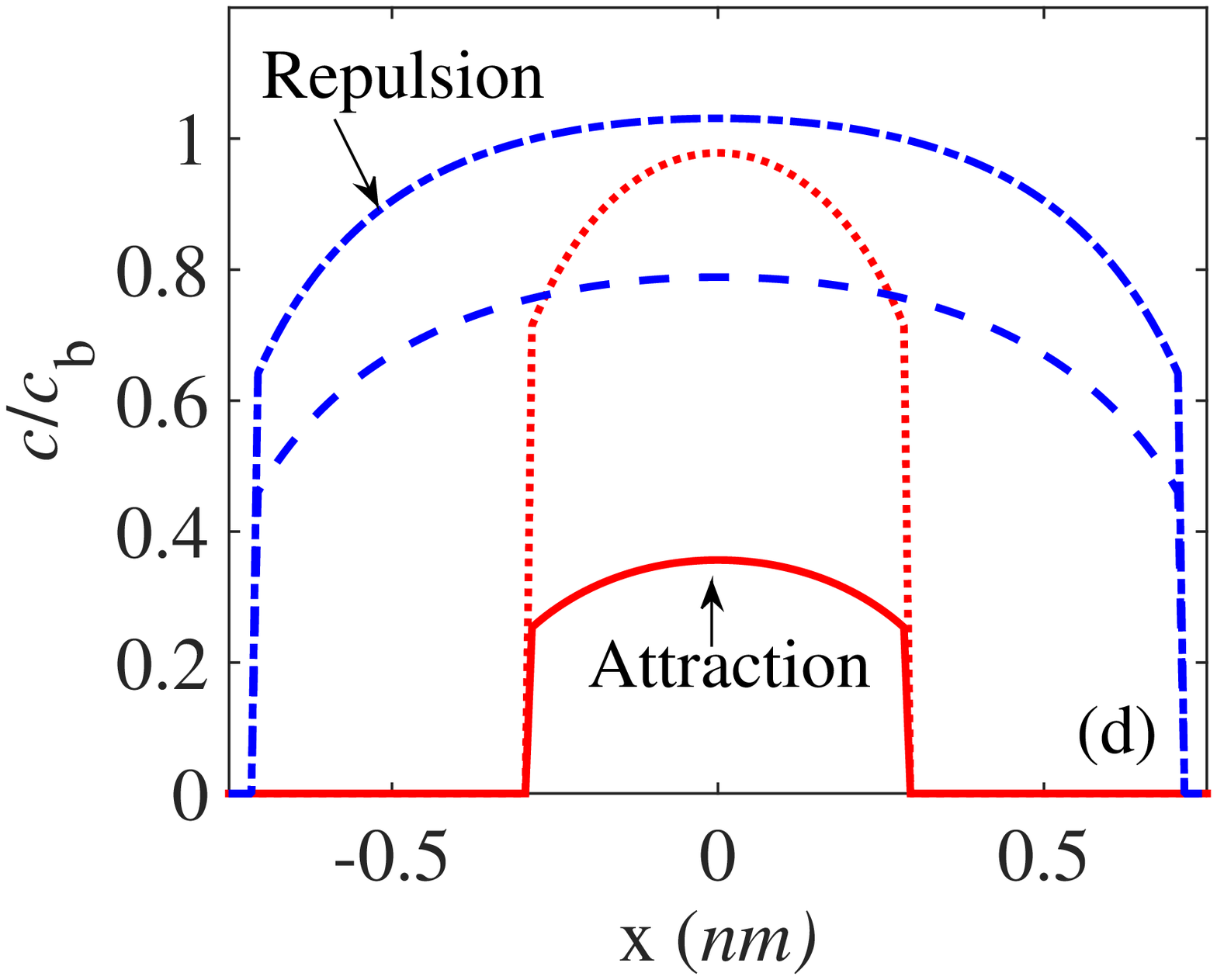}
\caption{Results for 1:1 electrolytes with uniform ionic sizes of radius $a_\pm=0.2 nm$. (a) Pressure of different surface charge densities as function of separation distance in an electrolyte of bulk concentration $c_{\pm}^b =0.1 M$;  (b)Pressure of planes with the same charge density $\sigma_1=\sigma_2=-0.01 e/nm^2$ in electrolytes of different bulk concentrations; (c) Phase diagram for the LCA boundary of surface-charge and salt-concentration with the same $\sigma$ on two planes; (d) Ionic densities when $D_\mathrm{eff}=0.58$ and $1.42 nm$ for systems of $c_{\pm}^b =0.1 M$ from Panel (b). } \label{ex1}
\end{center}
\end{figure*}

The pressure between surfaces depends on the difference in ionic densities between the bulk electrolyte and the mediated electrolyte. This is certainly relevant to the surface charge density $\sigma$ and bulk ionic number density. For symmetric electrolytes, the bulk ionic densities for both species are equal, $c_+^b=c_-^b=c_b$. Higher $\sigma$ increases the total ions in the mediated electrolyte, as a result the attraction between surfaces is reduced.  Fig. \ref{ex1} (a) and (b) show pressure curves as function of the effective separation for systems of 1:1 electrolytes with uniform ionic sizes, $a_\pm=0.2 nm$, where the abscissa represents the effective distance between two interfaces, $D_\mathrm{eff}=D-a_+-a_-$. The surface charge density takes $-0.01$ and $-0.015 e/nm^2$, corresponding to the coupling parameter $\Xi=0.032$ and $0.048$, respectively, in a weak-coupling regime.  For colloid-colloid interaction, one important characteristic length is the Debye length $\ell_D$, defined by $\ell_D^{-1}=\sqrt{\sum_i c_iz_i^2e^2/(\varepsilon_0\varepsilon_W k_BT)}$. For a $0.1 M$ electrolyte, it is $\ell_D=0.96 nm$. From Panel (a), the pressure remains positive for separation much smaller or much larger than the Debye length, and shows an attractive pressure when the separation distance is comparable to $\ell_D$, for which the attractive strength decreases with the increase of $\sigma$. With the decrease of the bulk salt concentration, the ionic density becomes smaller, reducing the entropic effect. This may reverse the attractive force at a certain concentration (shown in Panel (b)). We can see the attraction occurs when the distance is at the scale of the Debye length for the two denser cases, which is consistent with the particle-based simulation results \cite{Gan:JCP2015a}. The pressure does not depend much on the asymmetry in surface charge densities, but the strength of the densities. By this observation, we will only study planes with the same $\sigma$ in the following.

These above observations can be explained as that the pressure is resulted from the competition of the electrostatic repulsion and the entropically driven attraction \cite{CL:COCIS:2015}, and the LCA happens at the condition of smaller surface charge density for which the electrostatic repulsion is weak, or denser ionic density for which the entropically driven attraction is strong. This is in agreement with Wang and Wang \cite{WangRui:JCP:13} although they consider the ions as point charges and can not reach systems with higher surface charges due to the instability of the point-charge model \cite{XuMaggs:JCP:14}.  To study the behavior of higher surface charge density, we find the finite ion size is essential. The competition results in a phase boundary which separates the attractive region and the repulsive region in the $\sigma-c_b$ phase diagram, shown in
Fig. \ref{ex1} (c). We see at the  phase boundary, the relation is closes to $c_b \propto \sigma^2$, and accurate for small concentration up to $c_b\sim3 M$. The numerical results are shown for the bulk concentration up to $\sim 6 M$, which illustrate the planes can be attractive for surface charge about $0.1 e/nm^2$, corresponding to a system in the mediated-coupling regime. The electrolyte with concentration as high as $6 M$ is used in electrochemical experiments \cite{Ji:NC:2014a}.   This square law can be understandable since the electrostatic interaction is proportional to $\sigma^2$ at the weak-coupling regime, and the entropic contribution has the form of $c_b\log(c_b)$.

To a closer look at the mechanism, in Fig. \ref{ex1}(d), we plot the ionic densities for systems of $c_{\pm}^b =0.1 M$ and the surface charge densities $\sigma_1=\sigma_2=-0.01 e/nm^2$, corresponding to the solid circle line in Fig. \ref{ex1} (b) with separation $0.58$ and $1.42 nm$. It shows clearly that when the planes are attractive, the ionic densities are much smaller than the bulk densities, leading to a significantly less pressure due to the ionic depletion.

\subsection{Asymmetric valences and nonuniform sizes}

\begin{figure*}[t]
\begin{center}
\includegraphics[width=0.49\textwidth]{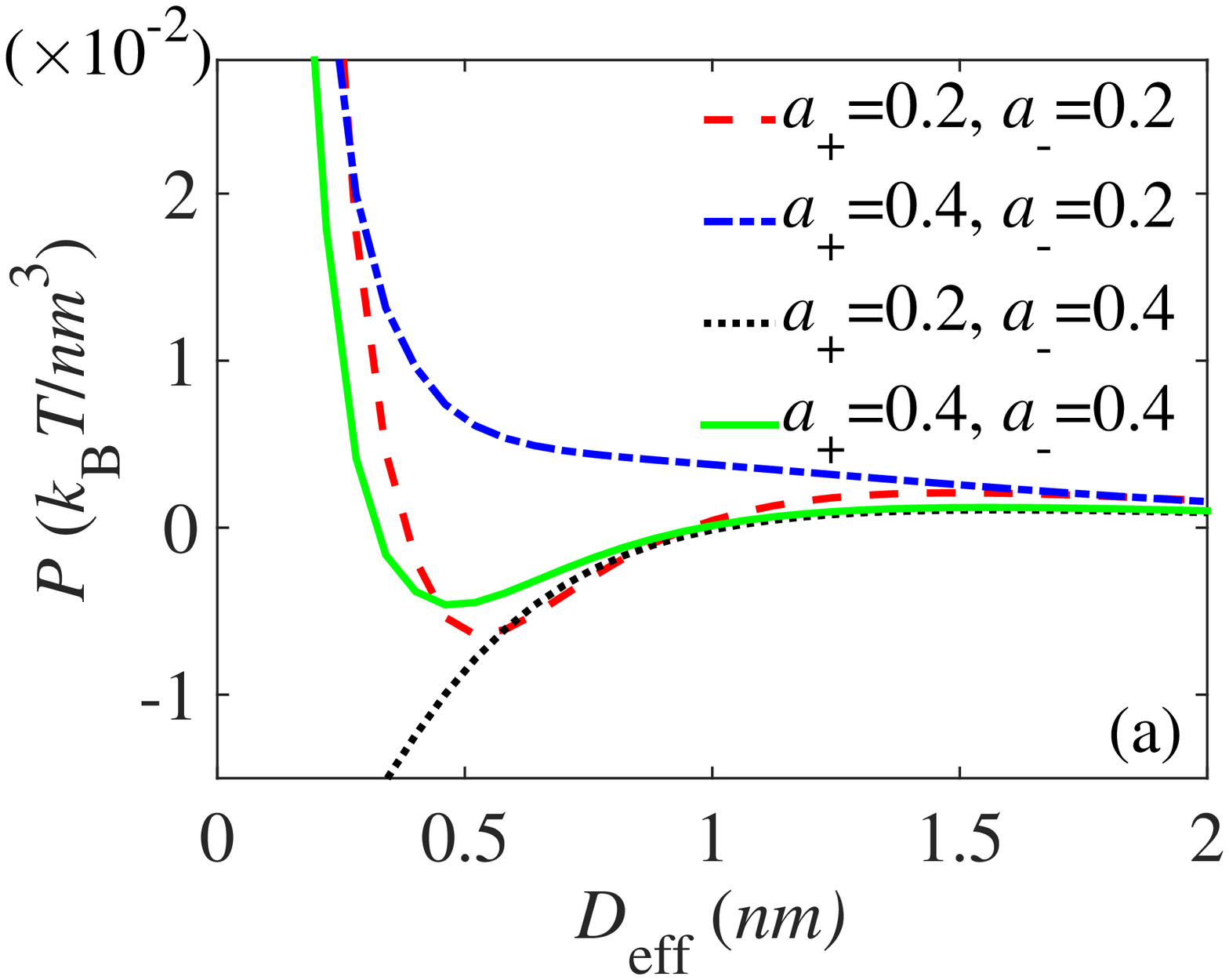} \includegraphics[width=0.49\textwidth]{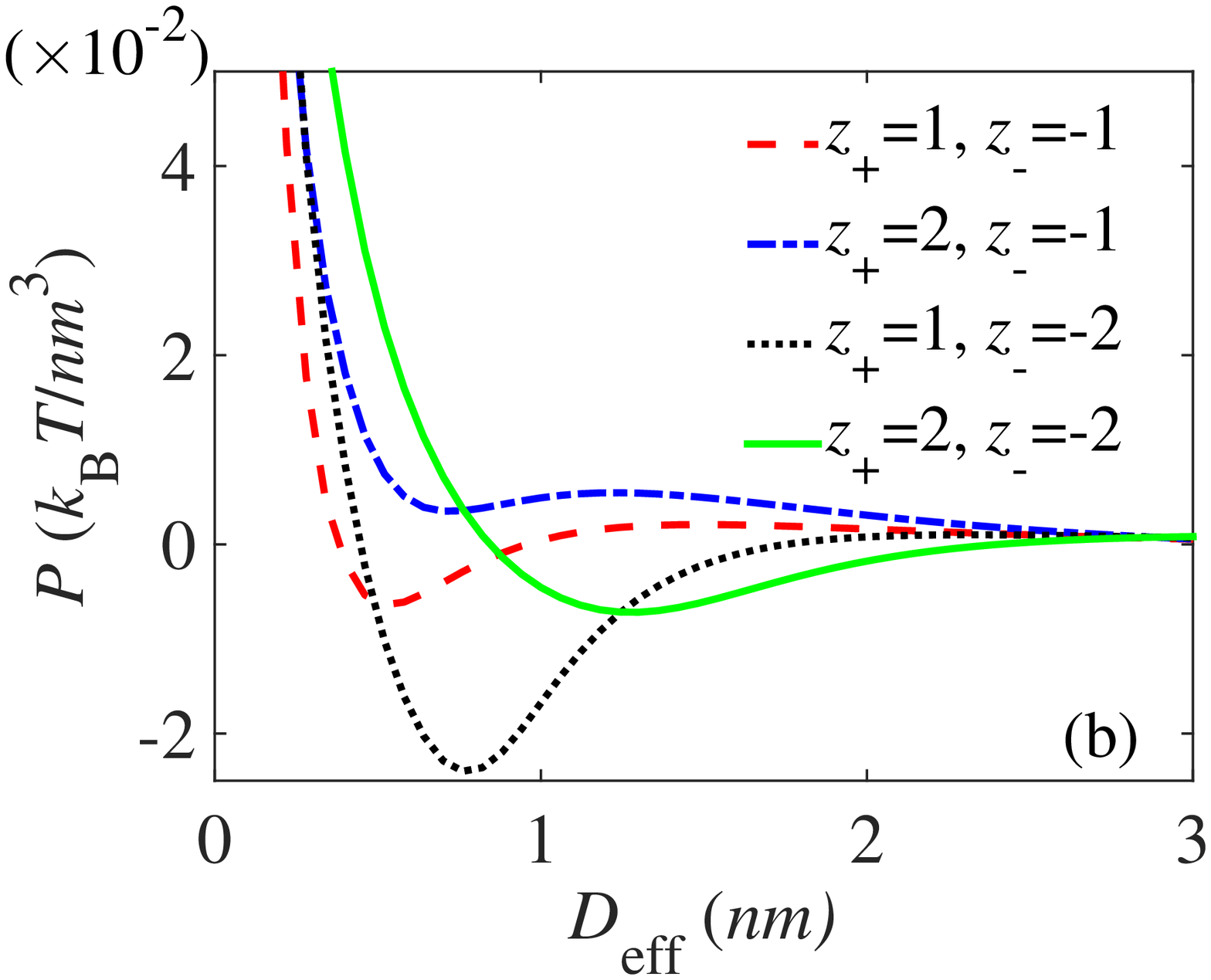}\\
\includegraphics[width=0.49\textwidth]{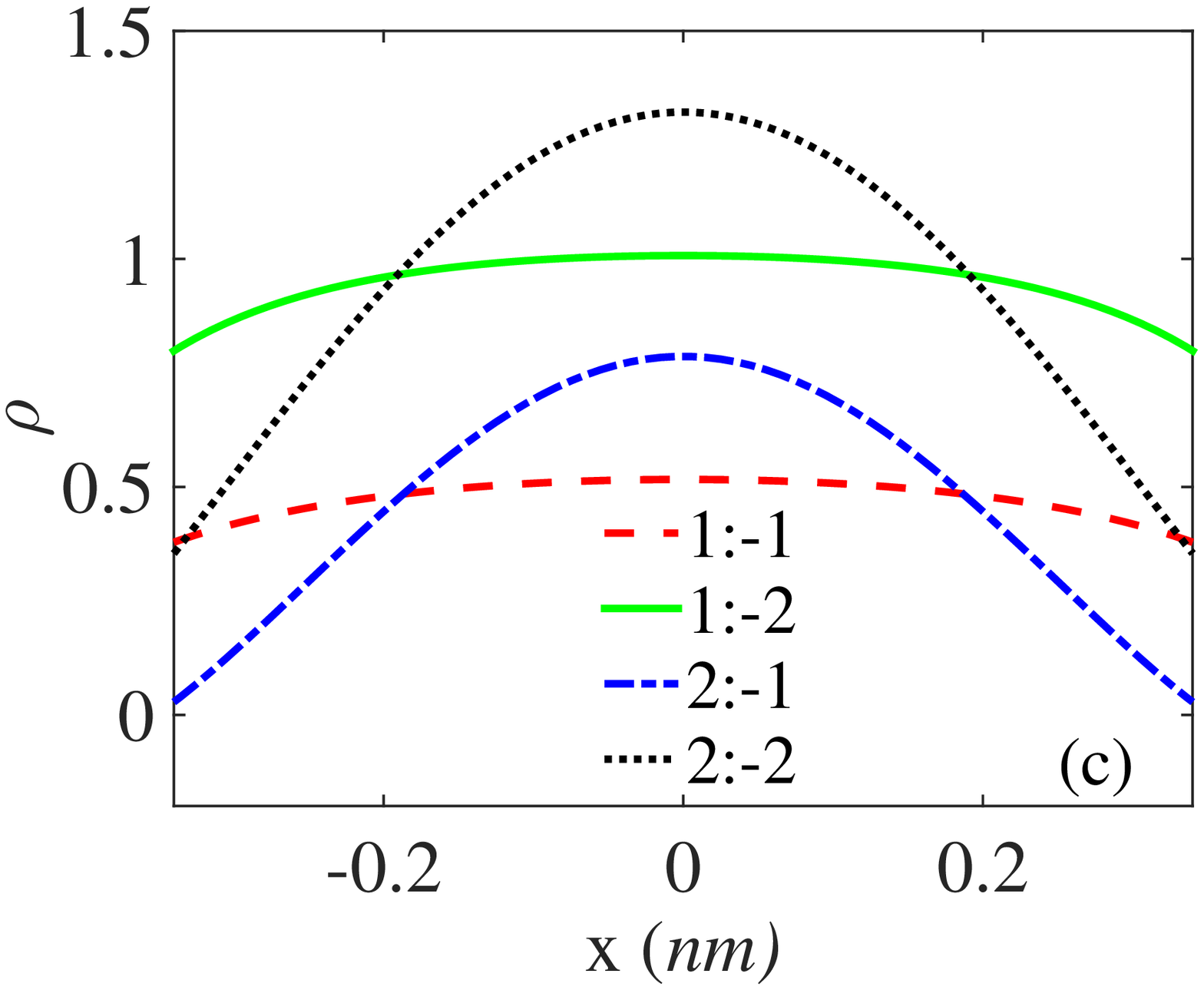} \includegraphics[width=0.49\textwidth]{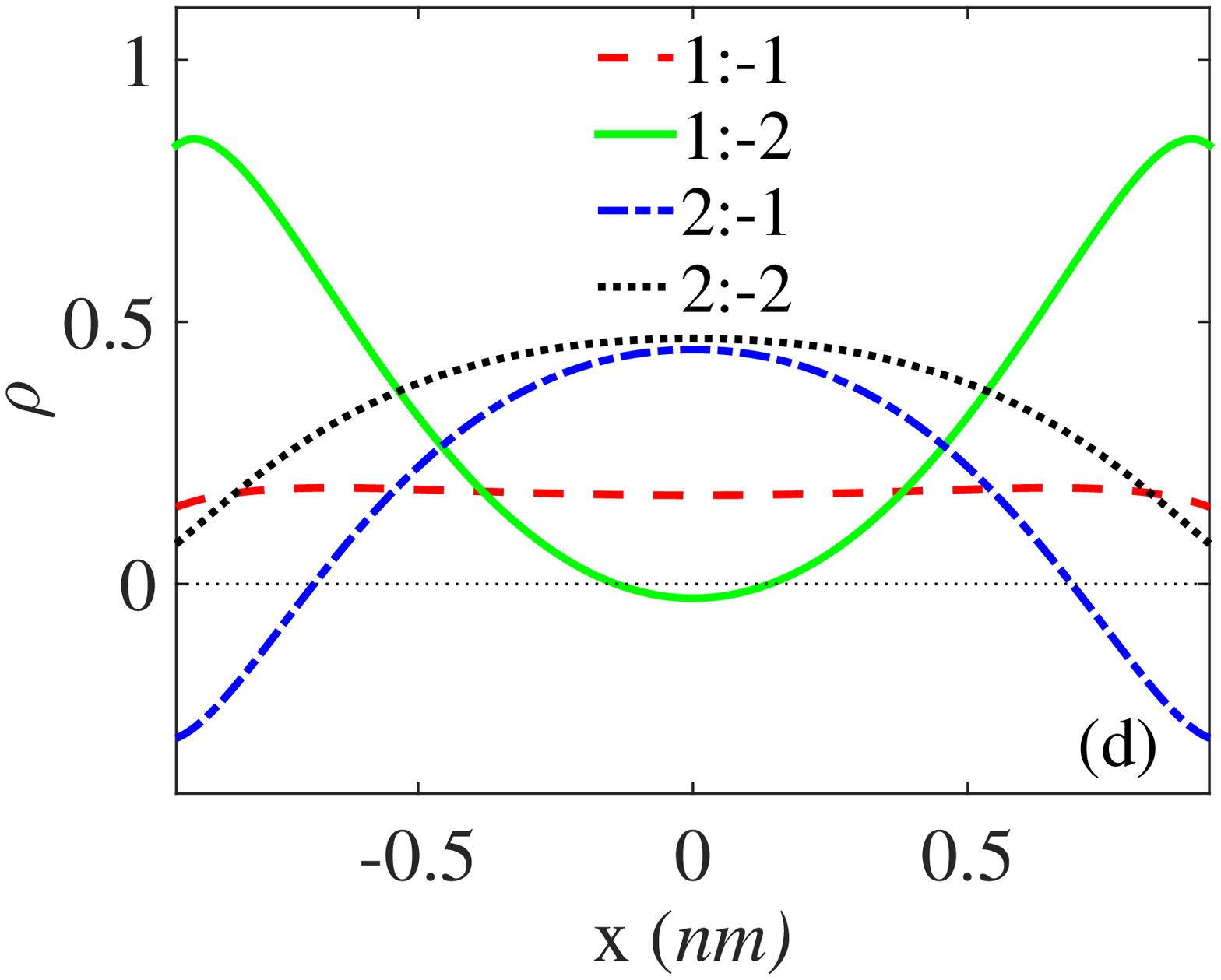}
\caption{Pressure and charge distribution for asymmetric electrolytes with $z_+c_+^b= 0.1 M$ between two planes of charge density $\sigma=-0.01 e/nm^2$. (a) Pressure between two planes for 1:1 electrolytes of variable ionic sizes;
(b) Pressure between two planes for asymmetric electrolytes and uniform ionic sizes;
(c)(d) Total charge density distribution along the $x$-axis in electrolytes for (c) $D_{\text{eff}}=0.7 nm$ and (d) $D_{\text{eff}}=1.9 nm$.
} \label{ex2}
\end{center}
\end{figure*}

The ionic size effect of ions is important for systems with nano-scale confinements. The inclusion of the ionic excluded volume and multivalent ions leads to a large change of the local correlation strength in self energy, and the consequence on the interface structure and interaction is not well investigated.  In this section, we perform numerical calculation for two groups of asymmetric systems. One group takes 1:1 electrolytes with variable ionic sizes; the other one takes uniform ionic sizes $a_\pm=0.2 nm$ but varies the ionic valences. The surface charge density and the bulk charge density remain constants, $\sigma=-0.01 e/nm^2$ and $z_+c_+= 0.1 M$. The results are shown in Fig. \ref{ex2}.

The ionic size matters in two aspects: 1) the thickness of the Helmholtz layer where ions are inaccessible; and 2) the self energy. From Fig. \ref{ex2}(a), we observe strong effects on pressure curves due to the non-uniformity of counterions and coions. For symmetric ionic sizes, the thickness of the Helmholtz layer is less important as we can see that the curves of $a_\pm=0.2 nm$ and $a_\pm=0.4 nm$ are similar while the other two curves is totally different.
For the effect of ionic size in the self energy, we should recall the formulation of the effective self energy,  the difference between $u_i$ and $u_i^b$,
\begin{equation}
u_i-u_i^b=\frac{u+\kappa_b/4\pi \varepsilon_0 \varepsilon}{1+ a_i \kappa_b},
\end{equation}
multiplied by the factor $z_i^2e^2/2$.
The leading contribution of $u$ is about $-\kappa(\mathbf{r})$ plus the boundary contribution including the image charge. If the ion is not too close to the surface, $u$ is in general a negative quantity, with$-u<\kappa_b$ due to the ion depletion. As a result, the self-energy contribution of a larger ion is less than that of a smaller ion, and hence the $a_\pm=0.4 nm$ curve behaves a weaker attraction than the $a_\pm=0.2 nm$ curve. If the ion size is very small which allows $u>0$, then the image charge effect dominates in this area. The image charge effect becomes weaker as the ionic size gets larger, again ensuring a weaker attraction.

For asymmetric ionic sizes, if counterions (cations) are bigger than coions (anions), the Helmholtz layer of counterions is thicker than coions'. Coions can be getting close to the charged surfaces, which effectively increases the density of the surface charge, thus the attractive force between interfaces is weakened or the interfaces become repulsive. This is shown by the curve of $a_-=0.2 nm$ and $a_+=0.4 nm$. In contrast, if coions are bigger, the surface charge density is effectively attenuated, leading to a stronger attraction potential (see the curve of $a_-=0.4 nm$ and $a_+=0.2 nm$). In this case, the planes of short separation remain attractive.

Let us look at the results in Fig. \ref{ex2}(b), the effects of valence asymmetry. Since the self energy has quadratic relation with the valence, $u_i \sim z_i^2$,  while ion-surface electrostatic force is roughly proportional to $z_i$, the multivalent counterion receives very strong repulsion from the surface. If coions have smaller valence, which cause less repulsion, equivalently the surface charge density is increased, thus the attractive force is weakened. The results have shown this situation: the planes in 2:1 electrolytes have weaker attraction (becomes pure repulsion) than those in 1:1 and 1:2 electrolytes.  The attractive distance also increases as the Debye length increases.

The panels (c) and (d) of Fig. \ref{ex2} present the total charge distribution along the normal direction of the interfaces for two separations $D_\mathrm{eff}=0.7 nm$ and $1.9 nm$ from panel (b). It can be observed that the 2:-1 electrolyte does show negative value near interfaces, validating the preceding analysis that the high-valence counterions increase the effective surface charge density. It should be mentioned that the curve for the 1:2 electrolyte in panel (d) is concave, different from the other three curves. Near the middle point, the total charge reverses the sign, illustrating a charge inversion, another important electrostatic many-body phenomenon \cite{GNS:RMP:2002,BKN+:PR:2005}. Again, it is owing to the strong repulsion of multivalent anions (counterions) by their image charge.

\subsection{Dielectric variation effect}

\begin{figure*}[t]
\begin{center}
\includegraphics[width=0.49\textwidth]{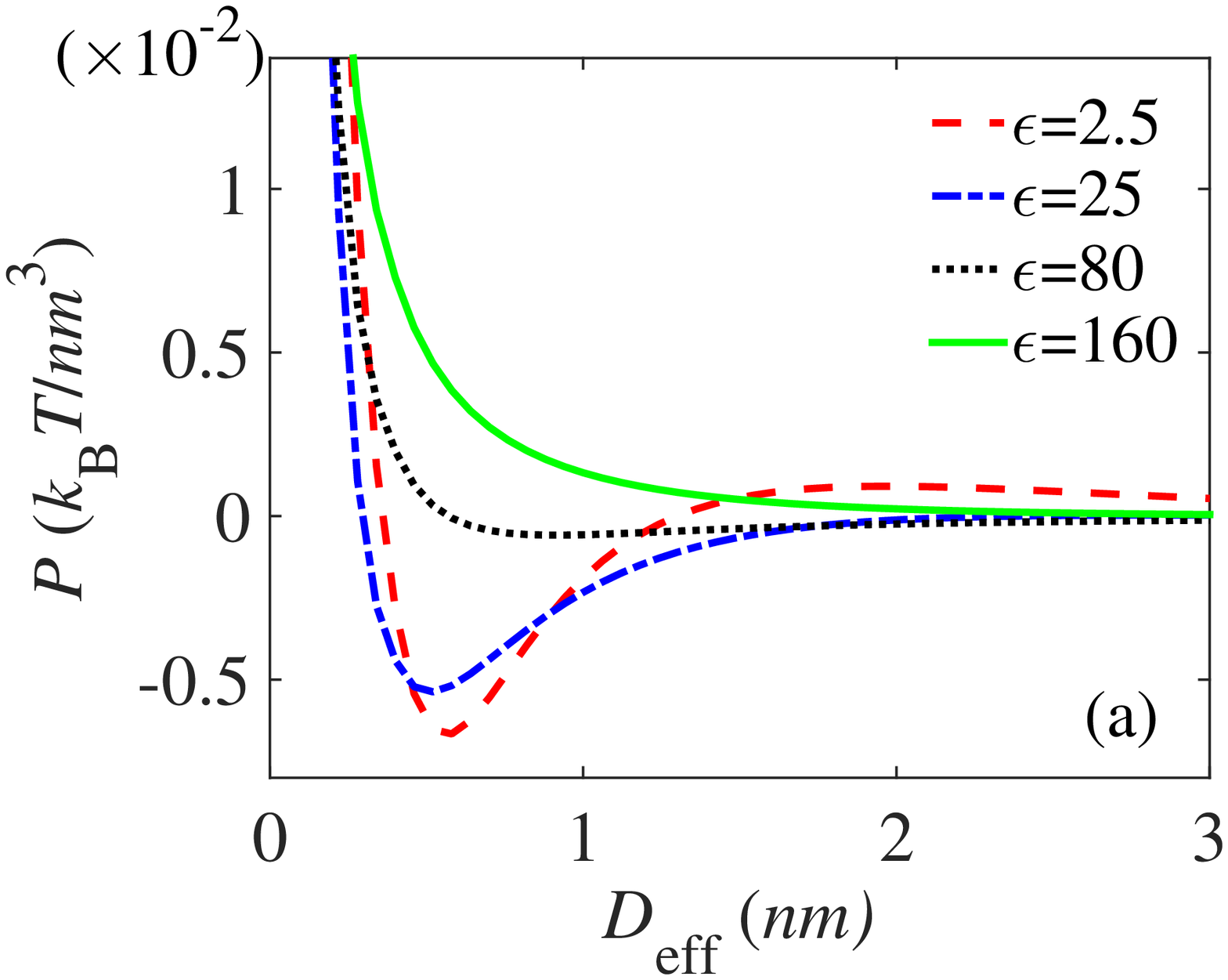} \includegraphics[width=0.49\textwidth]{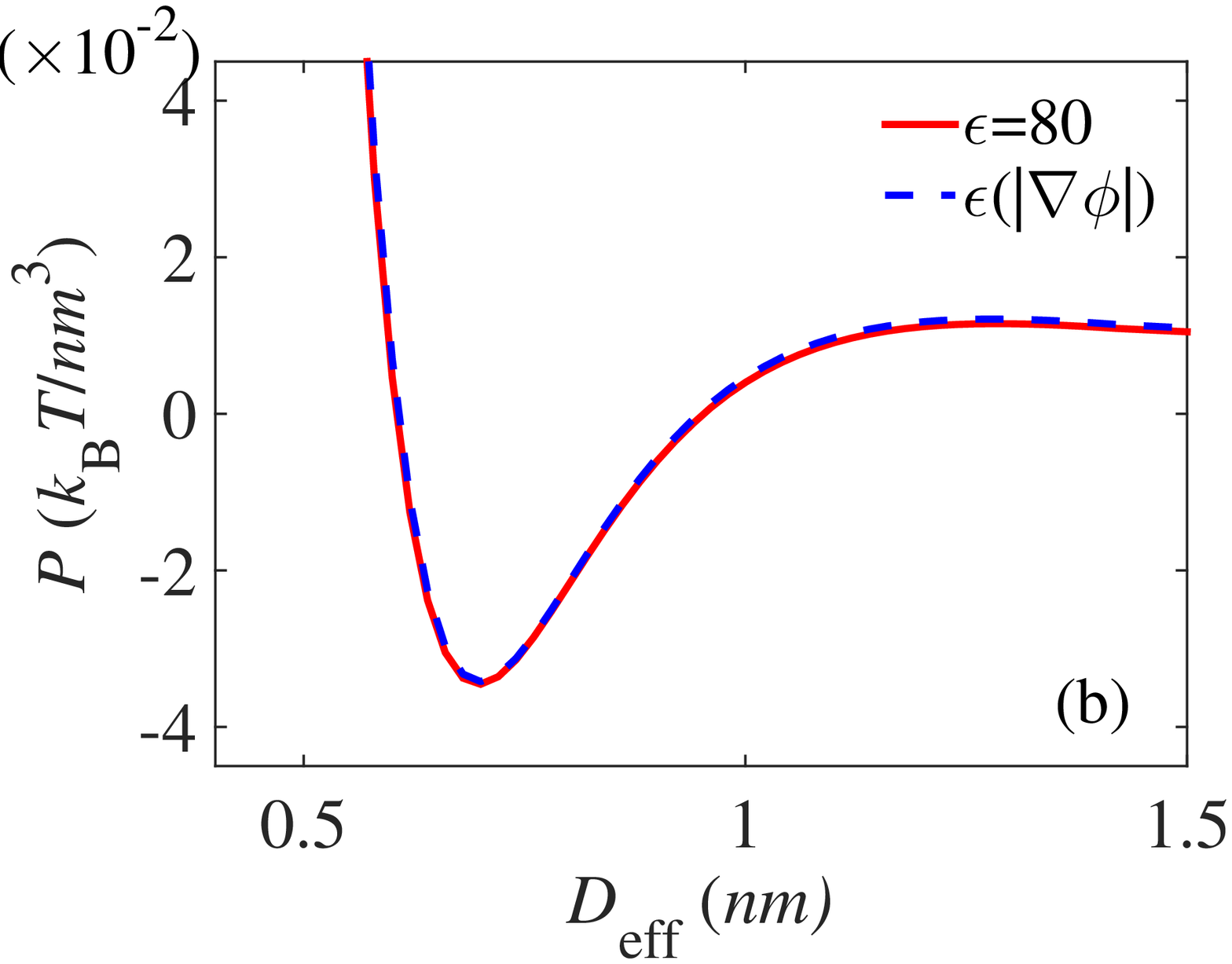}
\caption{Effects on dielectric permittivity on the pressure between two planes. The 1:1 electrolytes with $a_\pm=0.2 nm$  (a) Water dielectric constant $\varepsilon_W=80$ and varying boundary dielectric constant $\varepsilon_B$. Surface charge $\sigma=-0.01 e/nm^2$, bulk concentration $0.05 M$; (b) Boundary dielectric constant $\varepsilon_B=2.5$ and field-dependent dielectric permittivity for water. Surface charge $\sigma=-0.03e/nm^2$, concentration $0.5 M$. } \label{ex3}
\end{center}
\end{figure*}

We calculate the pressure between planes by varying the dielectric constant of the media outside the surfaces, $\varepsilon_B$, from 2.5 to 160. The results given in Fig. \ref{ex3}(a) illustrate a monotonic increase of the minimum pressure with $\varepsilon_B$. Compared to the homogeneous permittivity $\varepsilon_B=80$, the low dielectric constant of the boundary media greatly enhances the LCA. Interestingly, the planes become always repulsive when $\varepsilon_B$ is much larger than the water dielectric constant, e.g., $\varepsilon_B=160$. This shows a counter-intuitive phenomenon that the conducting limit of the planes tends to be repulsive due to the entropic forces.

In Fig. \ref{ex3}(b), we show the comparison with a variable permittivity in water. Since water molecule is polarizable, the dielectric permittivity of the water medium depends on the strength of the electrostatic field due to the orientation of dipole molecules against the field. A more ordered orientation leads to smaller permittivity, which can be described by the Langevin model \cite{FO:PCCP:11},
\begin{equation}
\varepsilon(|\nabla\Phi|)= 1 + 3(\varepsilon_W -1) \mathcal{L}(\beta p_0 |\nabla\Phi|)/ \beta p_0 |\nabla\Phi|,
\end{equation}
where $\mathcal{L}(y)=\coth(y)-1/y$ is the Langevin function, and $p_0$ is the permanent dipole moment. We take $p_0 = 4.8 D$ and $\varepsilon_W=80$. With the weak surface charge density, the results show the electric field near the surfaces slightly increases the interface-interface pressure which is due to a stronger electrostatic interaction in lower dielectric permittivity.

\section{Conclusions}

In summary, we study the phenomenon of the LCA between charged planes from a recently-developed self-consistent field model. We investigate the influence of different parameters of the electrical double layer on the many-body phenomenon. It is shown the depletion-induced LCA depends on the surface charge density and the bulk salt concentration, and the dielectric mismatch significantly changes the pressure strength between two planes. We find the mechanism of inducing LCA is mostly the depletion effect between the charged surfaces.
It is more likely to observe LCA for low surface charge, high salt concentration and the attraction can be greatly enhanced by small permittivity of the dielectric media outside the electrolyte. However, as the distance between charged surfaces which have LCA is comparable with Debye length, high salt concentration might lead to very narrow separation.

\section*{Acknowledgments}
The authors acknowledge the financial support from the Natural Science Foundation of China (Grant Numbers: 11101276, and 91130012), Youth Talents Program by Chinese Organization Department, and the HPC center of Shanghai Jiao Tong University.


\end{document}